# Understanding the local structure, magnetism and optical properties in layered compounds with $d^9$ ions: Insight into silver fluorides and $K_2CuF_4$

Inés Sánchez-Movellán, Guillermo Santamaría-Fernández, Pablo García-Fernández, José Antonio Aramburu* and Miguel Moreno

Departamento de Ciencias de la Tierra y Física de la Materia Condensada, Universidad de Cantabria, Cantabria Campus Internacional, 39005 Santander, Spain

* antonio.aramburu@unican.es

## Abstract

Using first-principles DFT calculations, we analyze the origin of the different crystal structures, optical and magnetic properties of two basic families of layered fluoride materials with formula $A_2MF_4$ (M = Ag, Cu, Ni, Mn; A = K, Cs, Rb). On one hand, $Cs_2AgF_4$ and $K_2CuF_4$ compounds (both with $d^9$ metal cations) crystallize in an orthorhombic structure with Cmca space group and $M_A$ - F - $M_B$ bridge angle of 180º, and they exhibit a weak ferromagnetism (FM) in the layer plane. On the other hand, $K_2NiF_4$ or $K_2MnF_4$ compounds (with $d^8$ and $d^5$ metal cations, respectively) have a tetragonal I4/mmm space group with 180º bridge angle and exhibit antiferromagnetism (AFM) in the layer plane. Firstly, we show that, contrary to what is claimed in the literature, the Cmca structure of $Cs_2AgF_4$ and $K_2CuF_4$ is not related to a cooperative Jahn-Teller effect among elongated $MF_6^{4-}$ units. Instead, first-principles calculations carried out in the I4/mmm *parent phase* of these two compounds show that $MF_6^{4-}$ units are axially compressed because the electrostatic potential from the rest of lattice ions force the hole to lie in the $3z^2 - r^2$ molecular orbital (z being perpendicular to the layer plane). This fact increases the metal-ligand distance in the layer plane and makes that covalency in the bridging ligand has a residual character (clearly smaller than in $K_2NiF_4$ or $KNiF_3$) stabilizing for only a few meV (7.9 meV for $Cs_2AgF_4$) an AFM order. However, this I4/mmm parent phase of $Cs_2AgF_4$ is unstable thus evolving towards the experimental Cmca structure with an energy gain of 140 meV, FM ordering and orthorhombic $MF_6^{4-}$ units. As a salient feature, it is shown that the FM order in $Cs_2AgF_4$ and $K_2CuF_4$ is due to the asymmetry of the in-plane $M_A$ - F - $M_B$ bridge, giving rise to a negligible covalency for the long bond. Moreover, in $K_2NiF_4$ or $K_2MnF_4$, the lack of excited states within the $d^n$ manifold (n = 8, 5) of M which can be coupled to the ground state for a local $b_{1g}$ distortion mode hamper the orthorhombic instability thus favoring the AFM ordering. The present ideas also account for the experimental optical and EPR data of $Cs_2AgF_4$ and $K_2CuF_4$. An additional discussion on the silver compound $Rb_2AgF_4$ is also reported.

## 1. Introduction

In the realm of insulating layered perovskites[1,2], much research is currently focused on those containing the 3d[9] ion $Cu^{2+}$ due to their attractive optical and magnetic properties[3-6]. Inorganic compounds[7-12] like $K_2CuF_4$ or $Rb_2CuCl_4$ as well as organic-inorganic hybrid perovskites of the $(C_nH_{2n+1}NH_3)_2CuCl_4$ or $[NH_3(CH_2)_nNH_3]CuX_4$ (X = Cl, Br) families[13-16] belong to this kind of materials. In all these insulating transition-metal compounds properties are greatly dependent on the equilibrium geometry and the related electronic structure of the $CuX_6^{4-}$ complex (X = F, Cl, Br) formed in the crystal.

In the last two decades significant efforts have been focused to exploring insulating compounds that contain the 4d[9] ion, $Ag^{2+}$, partially driven by the search of new superconducting materials[17]. As the optical electronegativity[18] of $Ag^{2+}$ ($\chi$ = 2.8) is higher than that of $Cu^{2+}$ ($\chi$ = 2.4) and identical to that of $Br^-$ ion[19], only crystals of silver(II) fluorides have been synthesized up to now. In chloride and bromide lattices $Ag^{2+}$ has been found only as impurity. For instance, $Ag^{2+}$ has been formed in alkali chlorides[20-22] and $SrCl_2$[20-22] while the $AgBr_6^{4-}$ complex has been observed in $CdBr_2:Ag^{2+}$ [23] and $AgBr_{0.15}Cl_{0.85}:Ag^{2+}$ [24] where the hole is mainly placed on ligands and not on the central cation[18,22].

Silver compounds of the series $A_2AgF_4$ (A = Cs, Rb) display a layered structure being $Cs_2AgF_4$ the most studied compound up to now[25-36]. From the first X-ray measurements, Odenthal et al. initially proposed[25] a tetragonal layered structure like that of $K_2NiF_4$ (space group I4/mmm) for $Cs_2AgF_4$. However, accurate neutron diffraction data by McLain et al. proved[28] that $Cs_2AgF_4$ is actually orthorhombic (space group Cmca) thus displaying the same crystal structure as $K_2CuF_4$.

In many previous works, the local geometry and electronic ground state of $AgF_6^{4-}$ units in $Cs_2AgF_4$ have been assumed to be the result of a Jahn-Teller (JT) effect and/or orbital ordering[26-36]. Nonetheless, the existence of a JT effect in $Cs_2AgF_4$ is hard to accept as the lattice is orthorhombic while doubts were already raised[11] on the orbital ordering as responsible for the structure of both $K_2CuF_4$ and $La_2CuO_4$. Moreover, we believe that the existence of a small ferromagnetic (FM) coupling in the layer plane of $Cs_2AgF_4$ and $K_2CuF_4$ has not been explained in a clear and convincing way. For example, no compelling reasons have been proposed to explain why the transition temperature, $T_c$, observed for both $Cs_2AgF_4$ (15 K)[28] and $K_2CuF_4$ (6.25 K)[7] is certainly smaller than that for the tetragonal perovskite $K_2NiF_4$ (97.3 K) or the cubic perovskite $KNiF_3$ (246 K) which are both antiferromagnetic[37,38] (AFM).

This work is firstly devoted to clearing up the origin of the equilibrium geometry and electronic ground state of $AgF_6^{4-}$ units in $A_2AgF_4$ (A = Cs, Rb) compounds paying special attention to $Cs_2AgF_4$. To achieve this goal, first-principles calculations together with an analysis of available experimental data are carried out. Once this key issue is clarified, we further discuss the origin of the FM ordering in the layer plane of $Cs_2AgF_4$ observed below $T_c$ = 15 K.

Interestingly, we try to verify in this work whether some new ideas used for understanding copper layered compounds[11,12,39] are also valid for the present case. Accordingly, we firstly focus on the I4/mmm parent phase of $Cs_2AgF_4$ paying attention to the internal electric field, $E_R(r)$, due to the rest of lattice ions on the behavior of the $AgF_6^{4-}$ unit where active electrons are confined. This internal field plays a key role for understanding the color of $Cr^{3+}$ gemstones[22,40] or the ground state[41,42] of $CuF_6^{4-}$ units in $K_2ZnF_4:Cu^{2+}$. For this reason, we explore here the influence of $E_R(r)$ on the nature of the electronic ground state of $AgF_6^{4-}$ complexes in $Cs_2AgF_4$ and the associated metal-ligand distances. If the parent phase's structure is different from the equilibrium geometry of $Cs_2AgF_4$, it implies the existence of an instability[12] discussed in a second step.

Available electron paramagnetic resonance (EPR) and optical data of $Rb_2AgF_4$ are quite close[26,43,44] to those found for $Cs_2AgF_4$ thus suggesting that $Rb_2AgF_4$ also exhibits a Cmca structure, a matter also analyzed in the present work.

This work is organized as follows: section 2 provides a detailed explanation of the computational tools employed. In section 3.1, an overlook on previous interpretations of structural, optical and magnetic data of layered silver fluorides is presented, while the remaining part of section 3 displays the main results obtained for $Cs_2AgF_4$ and $K_2CuF_4$. As $Cs_2AgF_4$ and $K_2CuF_4$ are model systems in the realm of layered compounds, section 3 is devoted to improving the knowledge on the structural (3.2), EPR (3.3) and optical data (3.4) of both compounds, as well as discussing the origin of their FM behavior not observed in $K_2NiF_4$ or $K_2MnF_4$ (3.5). Additionally, section 3.6 briefly discusses the main results obtained in $Rb_2AgF_4$. Finally, some conclusions are provided in section 4.

**2. Computational Tools**

First principles calculations based on Kohn-Sham DFT[45] were performed on pure $K_2CuF_4$, $Cs_2AgF_4$, $Cs_2CdF_4$, $Rb_2AgF_4$ and $K_2NiF_4$. In order to carry out periodic calculations, Crystal17 package was used[46,47]. This software makes a full use of the crystal symmetry, allowing us to perform geometry optimizations within specific space groups. In this sense, we can compute unstable phases, such as the high-symmetry parent phases of these compounds, that cannot experimentally be measured[11,12,48-50]. The instabilities occurring in these phases were explored through frequency calculations considering both FM and AFM orders, as discussed in section 3.5. In Crystal 17 code, Bloch wavefunctions are represented by linear combinations of Gaussian type basis functions[51,52] centred at atomic sites. The basis sets for the ions were taken from Crystal's webpage[52]. High-quality triple-zeta polarized basis developed by Peitinger et al. were employed[51]. For heavy ions ($Ag^{2+}$, $Cd^{2+}$ and $Cs^+$), pseudopotentials were used to describe the inner electrons. The exchange-correlation functional employed in these calculations was the hybrid PW1PW, which incorporates 20% of Hartree-Fock (HF) exact exchange.

This functional has been proven to yield accurate results for the geometry and properties of several pure and doped systems[53]. The obtained results were also cross-verified with another hybrid functional (B1WC) including 16% of HF exchange and different basis sets, leading to similar outcomes.

To ensure convergence, tight criteria were imposed for energy variations ($10^{-9}$ a.u.), as well as for root mean square (RMS) gradients and atomic displacement (0.0001 a.u.). A Monkhorst-Pack grid of 8x8x8 was used to sample the Brillouin zone for numerical integrations.

In addition to periodic calculations, cluster simulations were performed using the ADF software[54]. Hybrid B3LYP functional, which incorporates 25% of HF exchange, together with triple-zeta polarized basis sets were used in the calculations. The core electrons were kept frozen as they do not significantly influence the properties under study. These calculations make possible to explore optical spectra of both isolated and embedded complexes just considering[11,12,39,55,56] the potential created by the rest of lattice ions $V_R$. The electrostatic potential, particularly relevant in layered systems, was previously computed by Ewald-Evjen summations[57,58]. As shown in section 3.4, we can reproduce the experimental optical d-d transitions for $CuF_6^{4-}$ and $AgF_6^{4-}$ with reasonable accuracy when including the electrostatic potential $V_R$. Despite apparent differences in d-d transitions of these two systems, we demonstrate their consistent behaviour. Furthermore, our calculations on complexes provided valuable information about hybridization and covalency parameters of molecular orbitals, which are closely associated with the exchange constant J, as discussed in section 3.5.

**3. Results and discussion on $Cs_2AgF_4$ and $K_2CuF_4$**

*3.1 Survey of previous works on layered silver fluorides*

According to the results by McLain et al., $Cs_2AgF_4$ is a layered compound belonging to the orthorhombic Cmca space group[28], being **a** the crystal axis perpendicular to the layer plane (Fig. 1). The experimental values of lattice parameters and Ag-F distances are also shown on Fig. 1. $R_Z$ = 2.111 Å corresponds to the Ag-F direction perpendicular to the layer plane while $R_X$ = 2.112 Å and $R_Y$ = 2.441 Å are associated, respectively, with the short and long Ag-F distances within the layer plane.

$Cs_2AgF_4$ exhibits an antiferrodistortive arrangement and thus, if we move along the line joining two nearest $Ag^{2+}$ the local longest *y* axes of two involved $AgF_6^{4-}$ units, are perpendicular (Fig. 2). Interestingly, there is a weak FM exchange coupling, $H_{ex}$, between two nearest cations (with spins **S**$_1$ and **S**$_2$) in the layer plane of $Cs_2AgF_4$ described through the effective exchange interaction

$\qquad H_{ex} = J\ \mathbf{S}_1 \cdot \mathbf{S}_2$ \hfill (1)

with J = -3.79 meV[28]. This situation, also found in $K_2CuF_4$[7,28], is somewhat surprising as in layered compounds like $K_2NiF_4$ or $K_2MnF_4$ the coupling between magnetic cations in the layer plane is AFM[38].

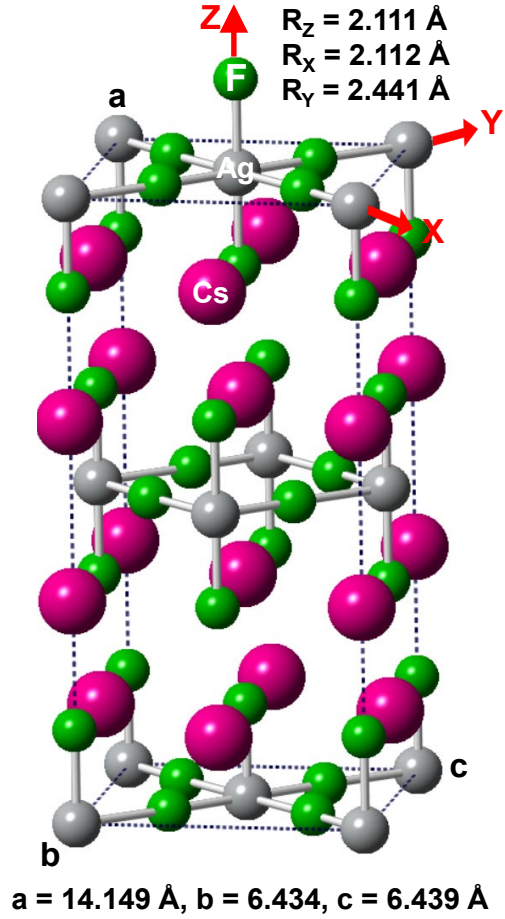

$R_Z$ = 2.111 Å
$R_X$ = 2.112 Å
$R_Y$ = 2.441 Å

a = 14.149 Å, b = 6.434, c = 6.439 Å

**Figure 1.** Experimental Cmca structure of $Cs_2AgF_4$ displaying the local {X,Y,Z} axes of a orthorhombic $AgF_6^{4-}$ unit. Values of the lattice parameters and the three $Ag^{2+}$-$F^-$ bond distances are shown.

Concerning EPR spectra of $Cs_2AgF_4$ in powder, two peaks are well observed[26,43]. One signal, g(**a**), corresponds to the magnetic field **B** parallel to the **a** axis while g($\perp$**a**) is the gyromagnetic factor when **B** lies in the layer plane. This isotropic response when **B** is in the layer plane obeys to the exchange interaction[59] between two nearest complexes, A and B. Indeed, using the local {x,y,z} bases of A and B complexes in Fig. 2, the g(i) (i = A, B) tensors in the crystal {X,Y,Z} basis are given by

$$g(A) = \begin{pmatrix} g_y & 0 & 0 \\ 0 & g_x & 0 \\ 0 & 0 & g_z \end{pmatrix} \quad g(B) = \begin{pmatrix} g_x & 0 & 0 \\ 0 & g_y & 0 \\ 0 & 0 & g_z \end{pmatrix} \quad (2)$$

Nevertheless, when the interaction of **B** with $S_1$ or $S_2$ is much smaller than $H_{ex}$ the phenomenon of exchange narrowing takes place[59] and then $g(\perp a)$ is just given by

$$g(\perp a) = (g_x + g_y)/2 \qquad (3)$$

If $J$ = -3.79 meV, the condition $|J| \gg \mu_B B$ ($\mu_B$ is the Bohr magneton) is well fulfilled working in the $Q$ band (B = 1.2 T) or even using B = 10 T[26].

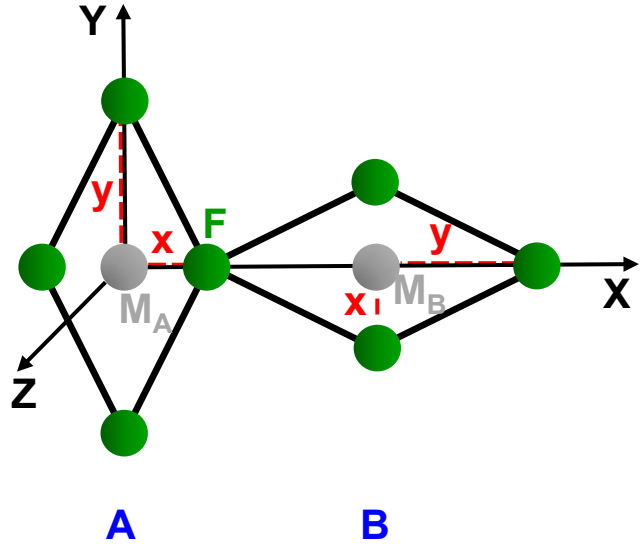

**Figure 2.** Schematic view of the in-plane antiferrodistortive arrangement of two adjacent A and B $AgF_6^{4-}$ units in $Cs_2AgF_4$, where the two local longest *y* axes are perpendicular.

The local geometry around $Ag^{2+}$ in $Cs_2AgF_4$ has widely been assumed to arise from a JT effect leading to elongated tetragonal $AgF_6^{4-}$ units[26,28,30,33,35,43] with *y* as the local principal axis of the complex and the hole placed in a molecular orbital transforming like $z^2 - x^2$. To underpin this view, McLain et al. argue[28] that for JT complexes of $d^9$ ions the hole is always placed in an orbital of the type $x^2 - y^2$. Accepting the JT assumption, the experimental $g(\mathbf{a}) = g_z$ value should then correspond to the $g_\perp$ quantity characteristic of a tetragonal complex while $g_y = g_\parallel$ and $g_x = g_\perp$. Therefore, in that case $g(\perp a)$ should be equal to

$$g(\perp a) = (g_\perp + g_\parallel)/2 \qquad (4)$$

Nonetheless, doubts are raised by the JT assumption due to the following reasons:

(i) The existence of a JT effect requires a degenerate electronic state in the *initial* geometry[60-62]. Even if that parent geometry corresponds to a I4/mmm tetragonal phase the electronic ground state of an $AgF_6^{4-}$ unit should not be degenerate due to symmetry reasons.

(ii) $Cs_2AgF_4$ is a layered compound where layers are perpendicular to the **a** crystal axis. Therefore, it is surprising that under the JT assumption the principal axis of the $AgF_6^{4-}$ unit is in the layer plane and not perpendicular to it.

(iii) The local equilibrium geometry for $AgF_6^{4-}$ in $Cs_2AgF_4$ is not tetragonal but orthorhombic although $R_Z$ and $R_X$ are very close as shown in Fig. 1.

(iv) From Eq. (4) and the experimental values g(**a**) = 2.07 and g($\perp$**a**) = 2.25 we get $g_\parallel$ - $g_0$ = 0.42 and $g_\perp$ - $g_0$ = 0.07 which are rather different from $g_\parallel$ - $g_0$ = 0.57 and $g_\perp$ - $g_0$ = 0.11 measured for the cubic lattice $CsCdF_3$ doped with $Ag^{2+}$, where EPR data prove the existence of a static JT effect with elongated $AgF_6^{4-}$ units[63].

(v) Optical spectra on $Cs_2AgF_4$ and $Rb_2AgF_4$ by Friebel and Reinen in the range 0.5 - 3.5 eV suggest the existence of four and not three *d-d* transitions[43], a fact consistent with a local orthorhombic symmetry around $Ag^{2+}$. However, in the case of $K_2CuF_4$, that also displays a Cmca structure, only three d-d transitions have been observed[10] even working at T = 5 K, a matter that is certainly puzzling[64].

(vi) Although most of $d^9$ systems which exhibit a static JT effect are elongated with a hole in a $x^2$ - $y^2$ type orbital[21,60,62] this is not a general rule. For instance, in the cubic CaO lattice doped with $Ni^+$ the hole is in a $3z^2$ - $r^2$ orbital and the octahedron compressed[65,66]. In non-JT systems like $K_2ZnF_4$:$Cu^{2+}$, where the host lattice is tetragonal[41,42], the hole is also in $3z^2$- $r^2$.

The following sections are addressed to clarify all these questions with the help of first principles simulations.

*3.2 Origin of the structure and electronic ground state*

In order to explain in a consistent way the experimental results on $Cs_2AgF_4$ exposed in section 3.1, it is crucial to know the actual ground state of involved $AgF_6^{4-}$ units and the origin of its equilibrium geometry. For achieving this goal, we have followed 3 steps:

(1) Firstly, we have determined the *parent phase* of the Cmca structure of $Cs_2AgF_4$ by substituting all $Ag^{2+}$ ions of $Cs_2AgF_4$ by the closed shell cation $Cd^{2+}$, with an ionic radius similar to that of $Ag^{2+}$, and performed a geometry optimization keeping fixed the Cmca space group. As shown in Table 1, under the $Ag^{2+} \rightarrow Cd^{2+}$ substitution, the structure evolves from the initial orthorhombic to a final one where a = b and $R_X$ = $R_Y$, associated to a tetragonal I4/mmm symmetry. In this parent structure, $R_Z$ < $R_X$ although $R_X$ – $R_Z$ is equal only to 0.044 Å. This I4/mmm parent phase, characteristic of compounds like $K_2NiF_4$ or $K_2ZnF_4$, is the same previously found[11,12] in $K_2CuF_4$.

(2) Then, it is useful to explore the behavior of $Cs_2AgF_4$ fixing the I4/mmm structure of the parent phase. Interestingly, the results of this step (2) (Table 1) show that in the I4/mmm structure the $AgF_6^{4-}$ units are tetragonally compressed ($R_z$ < $R_x$ = $R_y$) with a hole in the antibonding molecular orbital transforming like $3z^2$ - $r^2$. Therefore, in this step the principal axis of the AgF$_6^{4-}$ unit is perpendicular to the layer plane. As it also happens[11,12,41,42] for K$_2$ZnF$_4$:Cu$^{2+}$ or K$_2$CuF$_4$, this ground state is helped by the action of the internal electric field **E**$_R$(**r**) on active electrons confined in the AgF$_6^{4-}$ unit. Fig. 3 depicts the energy of an electron (-e charge) of an AgF$_6^{4-}$ unit under the electrostatic potential, V$_R$(**r**), generating the internal electric field, **E**$_R$(**r**), through **E**$_R$(**r**) = -∇V$_R$(**r**). It can be seen that -eV$_R$(**r**) raises the energy of an electron placed in $3z^2 - r^2$ while it lowers that of a $x^2 - y^2$ electron and thus this fact alone favors locating the hole in a $3z^2 - r^2$ orbital. This behavior of the -eV$_R$(**r**) energy is thus similar to that found in compounds like K$_2$BF$_4$ (B = Mg, Zn, Cu) which also exhibit a layered structure[11,12,41,42].

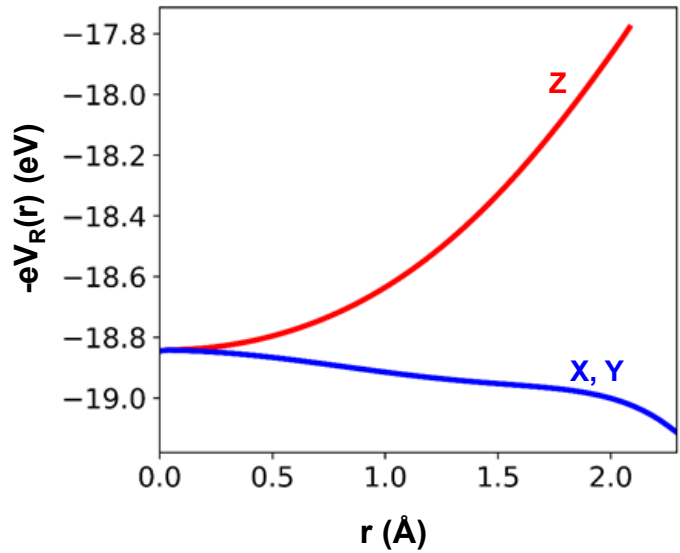

**Figure 3.** Potential energy –eV$_R$(**r**) corresponding to the internal electric field created by the rest of lattice ions on a AgF$_6^{4-}$ unit of Cs$_2$AgF$_4$ in the tetragonal I4/mmm parent phase, depicted along the local X, Y, and Z directions of the complex.

It is worth noting that on passing from Cs$_2$CdF$_4$ to Cs$_2$AgF$_4$ keeping the I4/mmm structure, R$_x$ – R$_z$ increases by a factor close to five. This significant difference just reflects that Ag$^{2+}$ has not a closed shell[58] structure as Cd$^{2+}$. Indeed, if in the tetragonal phase of Cs$_2$AgF$_4$ there is a hole in the $3z^2 - r^2$ orbital, this induces[67] an attractive force on two axial ligands and an increase of R$_x$ – R$_z$. Due to this process, at the end of the step (2) there are two Ag-F bonds with R$_Z$ = 2.081 Å while four *softer* at a distance R$_X$ = R$_Y$ = 2.292 Å.

(3) In a further step, when the I4/mmm restriction is lifted the calculated structure of Cs$_2$AgF$_4$ evolves until reaching a Cmca space group. As shown in Table 1, the optimized lattice parameters and Ag-F distances in step (3) coincide with those measured experimentally within 1.6%.

It should be remarked that, on passing from step (2) to step (3), $R_z$ undergoes a slight increase of 0.049 Å while $R_y - R_x$ moves from zero to 0.35 Å implying a significant distortion in the layer plane. This fact supports that the local equilibrium geometry of the $AgF_6^{4-}$ unit in the step (3) is essentially orthorhombic despite the experimental $R_x$ and $R_z$ values differ at the end by 0.001 Å (Table 1). Considering this fact alone, previous works on $Cs_2AgF_4$ have assumed[26,28,30,33,35,43] the existence of a JT effect with the local y as principal axis. The same assumption has been applied[9,10,68] to explain $K_2CuF_4$ where experimental $Cu^{2+}$-$F^-$ distances[8] are $R_z$ = 1.939 Å, $R_x$ =1.941 Å and $R_y$ = 2.234 Å.

| Step | compound | Space group | a | b | c | $R_z$ | $R_x$ | $R_y$ |
|---|---|---|---|---|---|---|---|---|
| (1) | $Cs_2CdF_4$ | Cmca | 14.773 | 6.392 | 6.392 | 2.216 | 2.260 | 2.260 |
| (2) | $Cs_2AgF_4$ | I4/mmm | 4.585 | 4.585 | 14.238 | 2.081 | 2.292 | 2.292 |
| (3) | $Cs_2AgF_4$ | Cmca | 14.382 | 6.454 | 6.455 | 2.130 | 2.106 | 2.458 |
| Experim. | $Cs_2AgF_4$ | Cmca | 14.149 | 6.434 | 6.439 | 2.112 | 2.111 | 2.441 |

**Table 1.** Step (1): calculated values of lattice parameters and Cd-F distances for $Cs_2CdF_4$ assuming an initial Cmca structure, which is the equilibrium structure of $Cs_2AgF_4$. It evolves until reaching an equilibrium tetragonal I4/mmm structure where b = c and $R_x$= $R_y$. Step (2): calculated lattice parameters and Ag-F distances corresponding to $Cs_2AgF_4$ but *fixing* a I4/mmm space group. Step (3): Idem that step (2) but derived for the Cmca equilibrium structure. The last values are compared to experimental ones[28]. Note that the c axis in the standard I4/mmm becomes the a axis in the standard Cmca. All distances are given in Å units.

The nature of the electronic ground state of $AgF_6^{4-}$ units derived from present calculations sheds light to clarify this key issue. Indeed, they reveal that the hole keeps a dominant $3z^2 - r^2$ character at the equilibrium geometry of $Cs_2AgF_4$ (step (3)). More precisely, an orthorhombic distortion allows the admixture of the molecular orbitals $|3z^2 - r^2\rangle$ and $|x^2 - y^2\rangle$ of the $AgF_6^{4-}$ unit for describing the wavefunction of the hole, $|\varphi_H\rangle$, and thus

$$|\varphi_H\rangle = \alpha|3z^2 - r^2\rangle - \beta|x^2 - y^2\rangle \qquad (5)$$

The present calculations give $\beta^2$ = 19% underlining that the orthorhombic distortion has only a moderate effect on $|\varphi_H\rangle$, a situation consistent with the degree of orthorhombicity given by the quantity $\eta = 2(R_y - R_x)/(R_y + R_x)$. Indeed, using the $R_y$ and $R_x$ values given in Table 3 we obtain $\eta$ = 14.5%.

The results in Table 1 support that the parent phase I4/mmm (where $R_y = R_x$) is not the stable one for $Cs_2AgF_4$. This instability is helped by the vibronic coupling between the electronic ground state of a $AgF_6^{4-}$ unit, $^2A_{1g}$, and $^2B_{1g}$ excited states driven by a $b_{1g}$ distortion mode described in Figure 4. That coupling is symmetry allowed and can lead to a total force constant that becomes negative (K < 0) thus favoring a Cmca phase at equilibrium. According to that distortion mode (Fig. 4) the four planar F$^-$ ligands of a given Ag$^{2+}$ forming initially a square give rise at the end to a rhombus where $R_Y \neq R_X$.

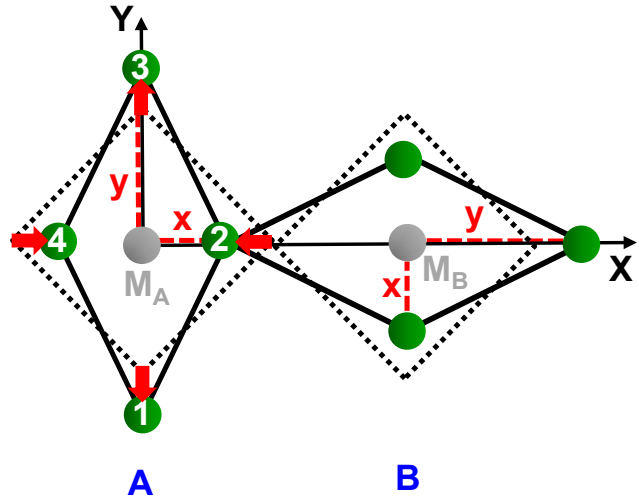

**Figure 4.** $b_{1g}$ distortion mode of the four planar F$^-$ ligands (red arrows) of a given AgF$_6^{4-}$ unit of Cs$_2$AgF$_4$, going from a square to a rhombus where $R_Y \neq R_X$. The $b_{1g}$ distortion mode can be described in terms of the normal coordinate Q, given by $Q = \frac{1}{2}(-\mathbf{u_1} - \mathbf{u_2} + \mathbf{u_3} + \mathbf{u_4})$, where $\mathbf{u_i}$ ($u_1 = -u_3 = u_2 = -u_4$) are the displacements of atoms 1-4.

To clarify this relevant matter a negative force constant can take place as there are two contributions[69] to the force constant K. Let us designate by H$_0$(**r**) the electronic Hamiltonian when the lattice is frozen at equilibrium positions. The corresponding ground state wavefunction is termed Ψ$_0$(**r**) while excited states are Ψ$_i$(**r**) with i > 0. If we move around the equilibrium position activating a Q coordinate a new vibronic term, h(Q,**r**), has to be added to the Hamiltonian

$$H = H_0(\mathbf{r}) + h(Q,\mathbf{r}) \qquad (6)$$

with h(0,**r**) = 0. A first elastic contribution to the force constant, called K$_0$, comes from considering only h(Q,**r**) and Ψ$_0$(**r**), thus neglecting the changes on the ground state wavefunction caused by the Q distortion. However, the presence of h(Q,**r**) also makes that the ground state wavefunction is no longer Ψ$_0$(**r**) but involves an admixture of excited states Ψ$_i$(**r**) giving rise to a negative vibronic contribution, K$_v$, to the force constant. If Q is a non-symmetric non-degenerate vibration mode transforming as irrep Γ, h(Q,**r**) is usually expanded as

$$h(Q,\mathbf{r}) = V(\mathbf{r})Q + \ldots \qquad (7)$$

where, in the linear electron-vibration coupling, V(**r**) transforms like Q. In second-order perturbation theory, this coupling give rise to the following expression for K$_v$

$$K_v = -\sum_{i>0} \frac{|\langle \Psi_0|V(r)|\Psi_i\rangle|^2}{E_i - E_0} \tag{8}$$

Accordingly, $K_v < 0$ for the electronic ground state, and K is just given by[69]

$$K = K_0 - |K_v| \tag{9}$$

The excited states in Eq. (8) must have the same spin as the ground state. Moreover, if $\Psi_0(r)$ and $\Psi_i(r)$ belong to the irreps $\Gamma_0$ and $\Gamma_i$, respectively, non-zero matrix elements in Eq. (8) are obtained only when $\Gamma_0 \otimes \Gamma_i \supset \Gamma$. Therefore, in a tetragonal $MF_6^{4-}$ complex (M = Cu, Ag), a $^2A_{1g}$ ground state can be mixed with an excited $^2B_{1g}$ state, involving a hole in $x^2 - y^2$, through the vibronic coupling involving a local $b_{1g}$ distortion mode[70].

According to Eq. (9), the instability develops when K becomes negative and is favored for low values of $K_0$ and thus soft bonds. If initially there is a compressed tetragonal complex, then $R_z < R_y = R_x$ and thus the equatorial bonds are softer than axial ones. This fact favors in principle the development of an orthorhombic distortion in the equatorial plane[11,12,39,67] provided K < 0. On the contrary, when the complex is elongated and $R_z > R_y = R_x$, it hardens the equatorial bonds a fact that works against the distortion. For this reason, in systems like NaCl:$M^{2+}$ [20,21,71] or $CsCdF_3$:$M^{2+}$ (M = Cu, Ag)[49] where $MX_6^{4-}$ units (X = Cl, F) are tetragonally elongated, no orthorhombic distortion in the equatorial plane has been observed. The same situation is encountered for $La_2CuO_4$[11].

In the case of layered compounds, the condition K < 0 can be better fulfilled for chlorides than for fluorides where bonds are stronger. For this reason, the orthorhombic distortion in the layer plane has already been observed[72] for $Cu^{2+}$-doped $(CH_3NH_3)_2CdCl_4$ but not for $K_2ZnF_4$:$Cu^{2+}$ where $CuF_6^{4-}$ units display a compressed tetragonal symmetry[73]. Nevertheless, if we move to the pure compound $K_2CuF_4$ two adjacent $CuF_6^{4-}$ complexes share a common ligand and this fact has been shown[12] to help K to become negative. This situation is also found in $Cs_2AgF_4$.

*3.3 Interpretation of EPR data of $Cs_2AgF_4$ and $K_2CuF_4$*

The orthorhombic instability taking place in $Cs_2AgF_4$ obeys to the same grounds previously discussed[11,12] for $K_2CuF_4$. For this reason, we have now to reinterpret the EPR data obtained[26,43] for $Cs_2AgF_4$ based on $AgF_6^{4-}$ units initially compressed along the z axis that undergo a moderate orthorhombic distortion ($\eta$ = 14.5%) in the layer plane. For a compressed tetragonal $AgF_6^{4-}$ unit, $g_z$ differs from $g_0$ only in third-order perturbations[74] and thus the tetragonal expression is

$$g_z(T) = g_0 - \Delta g^{(3)} \tag{10}$$

For $AgF_6^{4-}$, the estimated third-order correction is $\Delta g^{(3)} \approx 0.02$. In a further step, a moderate orthorhombicity induces an admixture of $|x^2 - y^2\rangle$ in $|3z^2 - r^2\rangle$ (Eq. (5)), thus giving rise to a contribution to $g_z$ in second-order perturbations. Accordingly, the octahedral expression of $g_z$ in the equilibrium situation can be estimated by[39,70,74]

$$g_z(O) = g_0 - \Delta g^{(3)} + \beta^2\, \delta g_z(x^2 - y^2) \tag{11}$$

Here, $\delta g_z(x^2 - y^2) \equiv g_z(x^2 - y^2) - g_0$ corresponds to a hole fully located in a pure $|x^2 - y^2\rangle$ orbital. Taking $\delta g_z(x^2 - y^2) \approx 0.55$ from EPR data[49] of $CsCdF_3:Ag^{2+}$ and $\beta^2 = 19\%$ from present calculations for $Cs_2AgF_4$, we finally obtain $g_z(O) = 2.086$, a figure reasonably close to the experimental value[26] $g_z = 2.077$.

Although for a compressed tetragonal $AgF_6^{4-}$ unit with z as main axis $g_y = g_x = g_\perp(T)$ the orthorhombic distortion in the layer plane leads finally to $g_y(O) \neq g_x(O)$. As the orthorhombicity is moderate ($\eta = 14.5\%$), the difference between $g_y(O)$ and $g_x(O)$ comes mainly from the hybridization of $|3z^2 - r^2\rangle$ and $|x^2 - y^2\rangle$ (Eq. (5)). In this case, the expressions of $g_y(O)$ and $g_x(O)$ are[39,70,74]

$$g_y(O) - g_0 = [g_\perp(T) - g_0]\left[1 + \frac{\beta}{\sqrt{3}}\right]^2$$

$$g_x(O) - g_0 = [g_\perp(T) - g_0]\left[1 - \frac{\beta}{\sqrt{3}}\right]^2 \tag{12}$$

Therefore, $g_y(O) - g_x(O)$ is given by

$$g_y(O) - g_x(O) = \frac{4\beta}{\sqrt{3}}[g_\perp(T) - g_0] \tag{13}$$

Due to the phenomenon of exchange narrowing[45], when the magnetic field is in the layer plane of $Cs_2AgF_4$ only an isotropic signal $g(\perp a) = (1/2)[g_x(O) + g_y(O)]$ is observed. From Eq. (12) that quantity can be expressed as follows

$$\frac{g_x(O) + g_y(O)}{2} - g_0 = [g_\perp(T) - g_0]\left[1 + \frac{\beta^2}{9}\right] \tag{14}$$

Therefore, from the experimental value[26,43] $g(\perp a) = 2.25$ and the calculated $\beta^2 = 19\%$ we derive from Eqs. (13) and (14) the values $g_y(O) = 2.38$ and $g_x(O) = 2.12$.

It is worth noting now that a similar situation is encountered in the layered compound $K_2CuF_4$ where experimental EPR data[7] yield $g(a) = 2.08$ and $g(\perp a) = 2.30$ while a value $\beta^2 = 17\%$ is obtained from calculations[11]. Interestingly, in the layered hybrid perovskite $(CH_3NH_3)_2CdCl_4$ doped with $Cu^{2+}$ there is also an orthorhombic distortion giving rise to the experimental values[72,12], measured at T = 10 K, $g_z = 2.05$, $g_y = 2.33$ and $g_x = 2.12$. This pattern is thus comparable to that estimated for $Cs_2AgF_4$. The fingerprint of the orthorhombic instability in the chlorine plane is also found in $NH_4Cl$ containing the $CuCl_4(H_2O)_2^{2-}$ radical where EPR data[75] at T = 4.2 K yield $g_z = 2.02$, $g_y = 2.41$ and $g_x = 2.18$. Obviously, in these cases involving impurities, there is no exchange narrowing but the average signal at $(1/2)(g_x + g_y)$ is also well seen upon raising the temperature due to a motional narrowing process[74,20,21,48].

*3.4 Interpretation of optical data of $Cs_2AgF_4$ and $K_2CuF_4$*

Concerning optical spectra, it is first necessary to understand why only three d-d transitions[10] have been reported experimentally for $K_2CuF_4$ while four[43] for $Cs_2AgF_4$, despite both compounds exhibit the same Cmca structure.

|  | $x^2 - y^2 \rightarrow 3z^2 - r^2$ | $xz \rightarrow 3z^2 - r^2$ | $yz \rightarrow 3z^2 - r^2$ | $xy \rightarrow 3z^2 - r^2$ |
|---|---|---|---|---|
| Isolated $CuF_6^{4-}$ | 0.77 | 1.13 | 1.32 | 1.31 |
| $CuF_6^{4-}$ under $V_R(r)$ | 1.04 | 1.17 | 1.42 | 1.51 |
| Experimental | 1.03 | 1.17 | 1.50 | 1.50 |

**Table 2.** Calculated energy values (in eV) of four d-d transitions for a $CuF_6^{4-}$ unit in $K_2CuF_4$ at the equilibrium Cmca structure where experimental $Cu^{2+}$-$F^-$ distances are $R_z$ = 1.939 Å, $R_x$ = 1.941 Å and $R_y$ = 2.234 Å. For the sake of clarity, the transitions are firstly derived ignoring the internal potential $V_R(r)$, whereas its influence upon the $CuF_6^{4-}$ unit is considered in a second step. Experimental values[10] for $K_2CuF_4$ are also reported for comparison. Due to the local orthorhombic symmetry of $CuF_6^{4-}$ units in $K_2CuF_4$ a transition termed as $x^2 - y^2 \rightarrow 3z^2 - r^2$ only reflects the dominant character of involved orbitals.

The calculated d-d transitions for $K_2CuF_4$ at equilibrium are gathered in Table 2. These results firstly show the importance of considering the effect of the internal electric field $E_R(r)$ on the electrons confined in the $CuF_6^{4-}$ unit, where the shape[12] of the associated electrostatic potential, $V_R(r)$, is similar to that depicted in Fig. 3 for an $AgF_6^{4-}$ unit in $Cs_2AgF_4$. Once $V_R(r)$ is incorporated into the calculation the two lowest d-d transitions essentially coincide with those observed experimentally[10] (Table 2). Interestingly the present calculations for $K_2CuF_4$ reveal that there are four different transitions although the two highest ones derived at 1.42 eV and 1.51 eV are only separated by 0.09 eV. The comparison of this figure with the bandwidth (~0.30 eV) measured at T = 5 K of the band peaked at 1.50 eV reasonably explains why the two highest d-d transitions are not resolved in the optical spectrum of $K_2CuF_4$ in agreement with a previous work[50]

|  |  | $x^2 - y^2 \rightarrow 3z^2 - r^2$ | $xz \rightarrow 3z^2 - r^2$ | $yz \rightarrow 3z^2 - r^2$ | $xy \rightarrow 3z^2 - r^2$ | CT |
|---|---|---|---|---|---|---|
| Experim. | Friebel et al [43] | 1.59 | 1.92 | 2.27 | 2.45 | – |
|  | Tong et al [44] | – | 1.91 | – | 2.48 | 4.59 |
| Calculated | Isolated $AgF_6^{4-}$ | 1.03 | 1.70 | 1.94 | 1.94 | 5.0 |
|  | $AgF_6^{4-}$ with $V_R(r)$ | 1.43 | 1.74 | 1.97 | 2.18 | 4.94 |

**Table 3.** Experimental d-d transitions for $Cs_2AgF_4$ compared to those derived from present calculations. The transitions are firstly calculated ignoring the $V_R(r)$ potential while its influence upon the $AgF_6^{4-}$ unit is considered in a second step. Due to the local orthorhombic symmetry of $AgF_6^{4-}$ units in $Cs_2AgF_4$ a transition termed as $x^2 - y^2 \rightarrow 3z^2 - r^2$ only reflects the dominant character of involved orbitals. The energy of the charge

transfer (CT) transition observed[44] by Tong et al. is also given and compared to the calculated value. Transition energies are given in eV.

Experimental optical data for $Cs_2AgF_4$ are collected in Table 3 together with the results of present calculations. As shown in Table 3, calculated d-d transitions follow the experimental pattern but, in this case, they underestimate the experimental energies by ~ 0.2 eV. Again, the results gathered in Table 3 stress the significant role played by the internal field $E_R(r)$, specially for understanding the value of the lowest $x^2 - y^2 \rightarrow 3z^2 - r^2$ transition and the 0.2 eV gap between the $yz \rightarrow 3z^2 - r^2$ and $xy \rightarrow 3z^2 - r^2$ transitions. At variance with what happens for $K_2CuF_4$, the existence of such a gap can well be inferred looking at the optical spectrum[43] of $Cs_2AgF_4$.

In view of the method followed for clarifying the structure of $Cs_2AgF_4$ in section 3.2, it is worth analyzing the evolution of d-d transitions on passing from the I4/mmm parent phase to the equilibrium Cmca and the influence that the internal field, $E_R(r)$, has upon them[50,57]. Results of present calculations are displayed in Fig. 5. In the initial I4/mmm phase, ignoring $E_R(r)$, the calculated d-d transitions are consistent with a local compressed tetragonal symmetry of the $AgF_6^{4-}$ unit with z as main axis. Accordingly, the $xz \rightarrow 3z^2 - r^2$ and $yz \rightarrow 3z^2 - r^2$ transitions are degenerate and the highest d-d transition is $xy \rightarrow 3z^2 - r^2$. The local orthorhombic distortion associated with the I4/mmm $\rightarrow$ Cmca step increases significantly the gap between the mainly $3z^2 - r^2$ and $x^2 - y^2$ levels from 0.64 eV to 1.03 eV as a result of the repulsion of these orbitals, both belonging to the $a_g$ irrep under $D_{2h}$ symmetry. At the same time, as the distortion makes that $R_y$ becomes the longest metal-ligand distance, the energy of two levels yz and xy decreases and the $yz \rightarrow 3z^2 - r^2$ and $xy \rightarrow 3z^2 - r^2$ transitions are found to be accidentally degenerate. Finally, the inclusion of the $V_R(r)$ potential tends to increase the energy along z direction (Fig. 3). For this reason, it also enhances the energy of the transition between the mainly $x^2 - y^2$ and $3z^2 - r^2$ orbitals that moves from 1.03 eV to 1.43 eV. In the same vein, the $xy \rightarrow 3z^2 - r^2$ transition increases from 1.94 eV to 2.18 eV. Due to this fact that transition is no longer degenerate with $yz \rightarrow 3z^2 - r^2$ nearly insensitive to the effects of $V_R(r)$.

The measurements[44] by Tong et al. on $Cs_2AgF_4$ also show a band peaked at 4.6 eV ascribed to a charge transfer transition of the $AgF_6^{4-}$ unit. In the case of fluoride complexes a charge transfer excitation in the optical range (up to 6.2 eV) cannot usually be observed due to the high electronegativity of fluorine ($\chi$ = 3.9). This hindrance however disappears for $AgF_6^{4-}$ complexes[76] due to the high optical electronegativity of $Ag^{2+}$ ($\chi$ = 2.8). Interestingly, in the case of the layered $(C_2H_5NH_3)_2CdCl_4$ compound doped with $Cu^{2+}$ the first allowed charge transfer transition is observed at 3.1 eV[5]. Thus, according to the optical electronegativities[19] of chlorine ($\chi$ = 3.0) and $Cu^{2+}$ ($\chi$ = 2.4), we can estimate that such a transition would occur around 4.9 eV for $Cs_2AgF_4$. Our calculations (Table 3) lead to a transition energy equal to 4.95 eV.

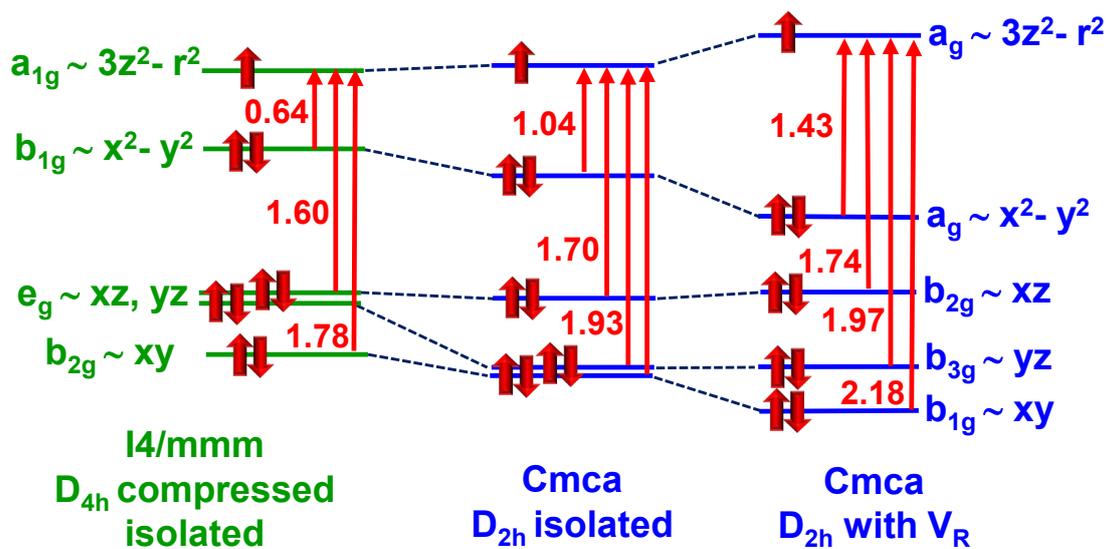

**Figure 5.** Evolution of mainly d orbitals of a $AgF_6^{4-}$ unit of $Cs_2AgF_4$ and the corresponding d-d transitions (given in eV units) along 3 steps. Left: isolated complex with the tetragonally compressed geometry ($D_{4h}$ symmetry with $R_y = R_x > R_z$) optimized for the metastable I4/mmm parent phase. Center: isolated complex with the orthorhombic geometry ($D_{2h}$ symmetry with $R_y \neq R_x$) optimized in the Cmca equilibrium structure. Right: same geometry than in previous step but adding the effect of the internal electrostatic potential $V_R(r)$.

*3.5 Insight into the ferromagnetism of $Cs_2AgF_4$ and $K_2CuF_4$ and comparison with $K_2NiF_4$*

This section addresses the origin of the two significant differences between the magnetic behavior of layered compounds $Cs_2AgF_4$ and $K_2CuF_4$ and that of the model system $K_2NiF_4$.

The main magnetic mechanism in play in these insulating lattices is superexchange that usually favors AFM ordering when the metal-ligand-metal bond occurs along a straight line. The main model describing superexchange, Anderson's model[77-79], is usually applied in a straightforward manner to $K_2NiF_4$. However, systems containing $d^9$ ions, like $Cs_2AgF_4$ and $K_2CuF_4$, are usually said to contain an orbital degree of freedom, i.e., the energy of the system and magnetic ordering are heavily influenced by how the occupied orbitals orient with respect to each other. Kugel and Khomskii proposed[80] the orbital ordering model based on two basic assumptions: (i) the $3z^2-r^2$ and $x^2-y^2$ orbitals are degenerate, and (ii) the energy of the system depends, essentially, on superexchange interactions between these orbitals. In 1973 they applied the orbital model to $K_2CuF_4$ predicting an antiferrodistortive orbital order at the I4/mmm phase, characterized by electrons in alternating $3x^2-r^2/3y^2-r^2$ orbitals, that would be the seed that leads to the orthorhombic distortion[81]. Moreover, they argued that this orbital antiferrodistortive order was necessary to obtain a FM ordering. However, the results in the previous section clearly show that the assumptions of the orbital ordering model are not correct. There exists a large gap between the $a_{1g}$ and $b_{1g}$ states in the I4/mmm parent phase so the orbitals are not degenerate favoring the hole to move to the $3z^2$-$r^2$ orbital in fluoride lattices. This FM ordering of the orbitals would be contrary to that predicted by the Kugel-Khomskii model and would lead to AFM ordering in the I4/mmm parent phase. This will be confirmed by first-principles simulations below. Moreover, if the orbital ordering is applicable to $K_2CuF_4$ it would also be applicable to $La_2CuO_4$, a system well known for its strong AFM ordering[11]. Thus, we believe that the use of the orbital ordering model in these lattices is compromised[6], and the origin of the magnetic structure should be re-analyzed in detail.

When in the $M_A$ - F - $M_B$ exchange path (Fig. 2) the involved angle, $\phi(M_A – F – M_B)$ is equal to 180º it usually gives rise to an AFM ordering, a situation found in the tetragonal $K_2NiF_4$ and $K_2NiF_4$ compounds belonging to the I4/mmm space group as well as in the cubic perovskite $KNiF_3$[38]. For this reason, the experimental FM arrangement displayed in the layer planes of both $Cs_2AgF_4$ and $K_2CuF_4$ compounds with a Cmca crystal structure and $\phi(M_A – F – M_B)$ = 180º is puzzling. Furthermore, there are also huge differences in the transition temperature, $T_c$. Indeed, $K_2NiF_4$ is AFM up to a temperature $T_c$ = 97.23 K[38], while for $K_2CuF_4$ and $Cs_2AgF_4$ the FM order disappears at much lower temperatures, namely at $T_c$ = 6.5 K [7] and $T_c$ = 15 K[28], respectively. This situation is, in principle, surprising as, in superexchange interactions, an increase of the covalency tends to enhance $T_c$ as it happens on passing from $RbMnF_3$ ($T_c$ = 83 K) to $KNiF_3$ (246 K)[38]. However, the optical electronegativity[19] of $Ni^{2+}$ ($\chi$ = 2.2) is smaller than that for $Cu^{2+}$ ($\chi$ = 2.4) or $Ag^{2+}$ ($\chi$ = 2.8)[18,76] and then, in a simple view, one would expect a smaller covalency and $T_c$ value for $K_2NiF_4$ than for compounds involving $d^9$ cations.

| Compound | Phase | Data | $R_z$ (Å) | $R_x = R_y$ (Å) | $\Delta E$ (meV) | J (meV) |
|---|---|---|---|---|---|---|
| $K_2NiF_4$ | I4/mmm | Experimental | 2.001 | 2.006 | 53.1 | 8.9 |
| $K_2CuF_4$ | I4/mmm | Calculated | 1.906 | 2.061 | 4.4 | 2.2 |
| $Cs_2AgF_4$ | I4/mmm | Calculated | 2.081 | 2.292 | 7.92 | 3.9 |

**Table 4.** Comparison of experimental data for the layered compound $K_2NiF_4$ [38], where the $Ni^{2+}$ spin is $S$ = 1, with those calculated in this work for $K_2CuF_4$ and $Cs_2AgF_4$ in the parent I4/mmm phase. The spin of $d^9$ ions is $S$ = ½. $\Delta E$ = E(FM) - E(AFM) per magnetic ion and depends on the exchange constant, J, and the spin of magnetic cation.

Seeking to clear these puzzling issues up we follow the procedure employed in section 3.2 thus exploring in a first step the behavior of $K_2CuF_4$ and $Cs_2AgF_4$ in the I4/mmm parent phase. To achieve this goal, we have calculated the total energy per magnetic ion, E(FM) and E(AFM), corresponding, respectively, to the FM and AFM phases associated to the same crystal structure and from them we derive the key quantity $\Delta E$ = E(FM) - E(AFM). It should firstly be stressed that the equilibrium geometry obtained for both magnetic configurations is essentially the same as they involve differences in metal-ligand distances smaller than $5\times10^{-3}$ Å. The obtained $\Delta E$ values for both $K_2CuF_4$ and $Cs_2AgF_4$ are displayed in Table 4 and lie in the range 1 – 10 meV and thus they are much smaller than |E(AFM)|. For instance, for $K_2CuF_4$ $\Delta E/|E(AFM)| \approx 5\times10^{-8}$, a fact that once more underlines the subtle origin of magnetic ordering.

The value $\Delta E$ = 53.1 meV for $K_2NiF_4$ in Table 4 is derived from the experimental exchange constant, J = 8.85 meV, associated with two coupled cations (Eq. (1)) using the expression

$$\Delta E = zJS_{max}(S_{max} + 1)/4 \qquad (15)$$

Here, z means the number of neighbor cations, equal to 4 for a layered compound, while $S_{max}$ is the maximum value of the total spin corresponding to two coupled cations with spin S, and thus equal to 2 for $K_2NiF_4$.

The main source of $\Delta E$ in compounds like $KNiF_3$ or $K_2NiF_4$ has been explained by Anderson and Hay et al. just considering a simple M - F - M symmetric dimer[77,79]. According to these models, $\Delta E$ is essentially related to *valence* electrons and involves the difference of two contributions

$$\Delta E = E_V(FM) - E_V(AFM) \qquad (16)$$

where the $E_V(FM)$ and $E_V(AFM)$ contributions associated only with valence electrons are both negative and favor the FM and AFM configuration, respectively. These contributions typically lie in the range 1 - 100 meV. In the same vein the exchange constant, J, of two coupled cations (Eqs. (1) and (15)) can be expressed as the difference between two positive contributions

$$J = J(AFM) - J(FM) \qquad (17)$$

According to the work by Anderson and Hay el al. on a symmetric dimer, the J(AFM) contribution greatly depends on the charge transferred from the cation to the ligand in orbitals containing *unpaired* electrons[77-79]. For clarifying this issue, we start considering a $NiF_6^{4-}$ unit in the cubic $KNiF_3$ compound that involves two unpaired electrons in the $|\theta\rangle$ and $|\varepsilon\rangle$ antibonding molecular orbitals (Fig. 6), described by[82]

$$|\theta\rangle = N\{|d(3z^2 - r^2)\rangle - (\mu/\sqrt{12})[2(|p_\sigma(5)\rangle + |p_\sigma(6)\rangle) - (|p_\sigma(1)\rangle + |p_\sigma(2)\rangle + |p_\sigma(3)\rangle + |p_\sigma(4)\rangle)]\}$$

$$|\varepsilon\rangle = N\{|d(x^2 - y^2)\rangle - (\mu/2)[|p_\sigma(1)\rangle - |p_\sigma(2)\rangle + |p_\sigma(3)\rangle - |p_\sigma(4)\rangle]\} \qquad (18)$$

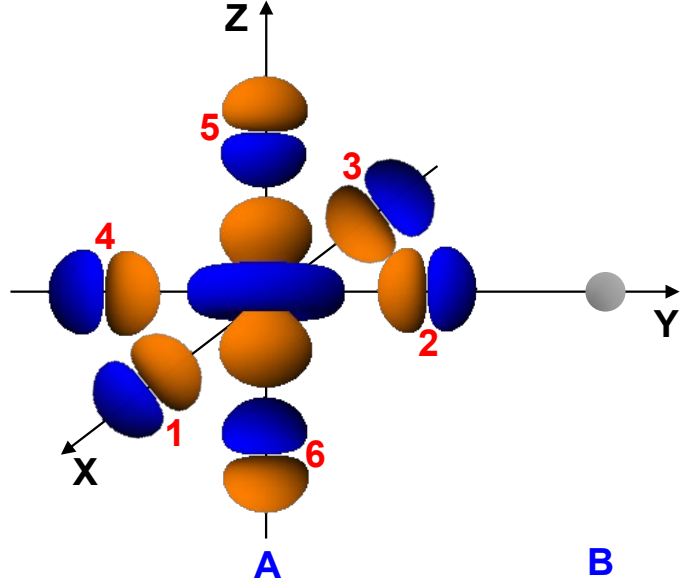

**Figure 6.** Picture of the antibonding $|\theta\rangle \sim 3z^2 - r^2$ molecular orbital for an octahedral $NiF_6^{4-}$ unit in cubic $KNiF_3$ perovskite.

Due to the existence of covalency, the two non-filled orbitals of a simple $NiF_6^{4-}$ unit have not pure $|d(3z^2- r^2)\rangle$ or $|d(x^2- y^2)\rangle$ characters thus involving a partial transfer of charge from metal to six ligands. If we consider the bridging ligand number 2 (Fig. 6), the charge transferred to it in σ-orbitals $|\theta\rangle$ and $|\varepsilon\rangle$ is equal to $(N\mu)^2/12$ and $(N\mu)^2/4$, respectively, thus implying a total charge, termed $f_\sigma$, equal to

$$f_\sigma(NiF_6^{4-}) = (N\mu)^2/3 \qquad (19)$$

Due to cubic symmetry in $KNiF_3$ the same charge is transferred to any ligand of the $NiF_6^{4-}$ unit. Obviously, if we consider a $M_A$ - F(2) - $M_B$ dimer (Fig. 6), the charge transferred from $M_A$ to the bridging ligand, $f_\sigma(A)$, is the same to that from $M_B$, $f_\sigma(B)$, provided the dimer is symmetric and then we can write $f_\sigma(A) = f_\sigma(B) = f_\sigma$. This symmetric situation happens in both $KNiF_3$ and $K_2NiF_4$ although in the last case only the four ligands[38] of a $NiF_6^{4-}$ unit in the layer plane are involved in the superexchange.

A central issue in the work by Anderson and Hay el al. on a symmetric dimer is the dependence of the J(AFM) contribution on $f_\sigma$ that, for the present purposes, can simply be written as[77-79]

$$J(AFM) \propto f_\sigma^2 \qquad (20)$$

This fact agrees with the measured decrease[38] of J on passing from $KNiF_3$ (8.4 meV) to $RbMnF_3$ (0.57 meV) following the smaller covalency[78, 82-85] in $MnF_6^{4-}$ ($f_\sigma \sim 1\%$) than in $NiF_6^{4-}$ ($f_\sigma \sim 4\%$), a result also consistent with the lower optical electronegativity of $Mn^{2+}$ ($\chi \approx 1.4$) [19,22,86,87] when compared[19] to $\chi = 2.2$ for $Ni^{2+}$.

The NiF$_6^{4-}$ complex in the tetragonal K$_2$NiF$_4$ compound is nearly octahedral (Table 4), with metal-ligand distances R$_x$ = R$_y$ = 2.006 Å and R$_z$ = 2.001 Å which are very close to R = 2.007 Å measured[38] in the cubic perovskite KNiF$_3$. As the Ni$^{2+}$ spin in the electronic ground state of K$_2$NiF$_4$ is also $S$ = 1 the exchange constant J in K$_2$NiF$_4$ (8.9 meV) and KNiF$_3$ (8.7 meV) derived from experimental data are very near. Despite this fact, the transition temperature T$_c$ is higher for KNiF$_3$ (246 K) than for the layered compound K$_2$NiF$_4$ (97.2 K) as a result of the different number of neighbor cations, z, in both compounds[88].

It is worth noting now that, if we move from K$_2$NiF$_4$ to K$_2$CuF$_4$ keeping the I4/mmm structure, the number of unpaired electrons is reduced from 2 to 1. Furthermore, the unpaired electron in K$_2$CuF$_4$ is located in a $|3z^2 - r^2\rangle$ molecular orbital of a D$_{4h}$ CuF$_6^{4-}$ complex helped by the electrostatic potential, V$_R$(**r**), due to the rest of lattice ions (Fig. 2). It should be noticed that in this orbital, coming from $|\theta\rangle$ in O$_h$ symmetry, the dominant bonding is with the two axial ligands, while the charge transferred to the in-plane bridging ligands is smaller. More precisely, if, as a rough first step, we approximate $|3z^2 - r^2\rangle$ by $|\theta\rangle$ we obtain for the bridge ligand f$_\sigma$(CuF$_6^{4-}$) = (Nµ)$^2$/12, which is thus four times smaller than the value f$_\sigma$ = (Nµ)$^2$/3 derived for (NiF$_6^{4}$) (Eq. (19)), provided Nµ is the same for both systems.

Seeking to clarify this central issue we have calculated the f$_\sigma$ value for both the bridging and axial ligands of K$_2$CuF$_4$ and Cs$_2$AgF$_4$ in the I4/mmm parent phase. Results are collected in Table 5 and compared to those for K$_2$NiF$_4$. They confirm that f$_\sigma$ is much higher for axial than for the in-plane bridging ligands of K$_2$CuF$_4$ and Cs$_2$AgF$_4$ thus stressing that covalency in the latter ligands is residual. At the same time, they show that AgF$_6^{4-}$ is more covalent than CuF$_6^{4-}$ in accord with the electronegativity scale. Despite these facts the f$_\sigma$ values derived for bridging ligands in K$_2$CuF$_4$ (1.70%) and Cs$_2$AgF$_4$ (2.43%) are clearly smaller than f$_\sigma$ = 4.1% calculated for K$_2$NiF$_4$ and KNiF$_3$. Values of f$_\sigma$ in the range 3.5% - 6% have been reported for NiF$_6^{4-}$ by other authors[78,82-85]. This relevant fact is thus consistent with the smaller J value derived for K$_2$CuF$_4$ and Cs$_2$AgF$_4$ when compared to that for K$_2$NiF$_4$ (Table 4). Now, if we take into account the different value of S$_{max}$(S$_{max}$ + 1) in both compounds, this explains why ΔE is one order of magnitude higher in the case of K$_2$NiF$_4$.

| Compound | Phase | f$_\sigma$(axial) (%) | f$_\sigma$(bridging) (%) | ΔE (meV) |
|---|---|---|---|---|
| K$_2$CuF$_4$ | I4/mmm | 7.98 | 1.70 | 4.4 |
| Cs$_2$AgF$_4$ | I4/mmm | 11.28 | 2.43 | 7.92 |
| K$_2$NiF$_4$ | I4/mmm | 4.1 | 4.1 | 53.1 |

**Table 5.** Calculated charge transferred to axial and equatorial ligands for the hole of MF$_6^{4-}$ units (M = Cu, Ag) in K$_2$CuF$_4$ and Cs$_2$AgF$_4$ compounds in the I4/mmm parent phase. Results are compared to those for K$_2$NiF$_4$ with two unpaired electrons. The ΔE values calculated for the three systems are also included for comparison.

A key question in the present analysis is to understand the influence of the I4/mmm → Cmca structural transition upon the magnetic ordering[6] of $K_2CuF_4$ and $Cs_2AgF_4$. In that process the $MF_6^{4-}$ complexes (M = Cu, Ag) change from tetragonal to orthorhombic symmetry a distortion described by the Q coordinate.

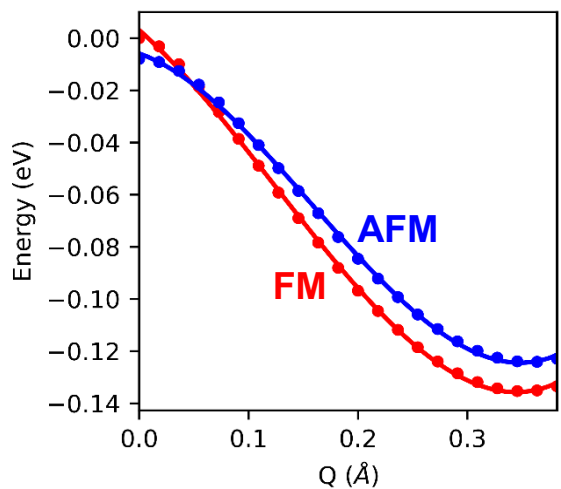

**Figure 7.** Energy (given per formula unit) of $Cs_2AgF_4$ obtained by single point calculations throughout the antiferrodistortive distortion from the parent tetragonal I4/mmm phase to the minimum at the orthorhombic Cmca one. These data are similar to those reported in Reference [6] although in that article the abscissa reflects the position of the bridge ligand and not the distortion coordinate, Q.

The variation of ΔE = E(FM) - E(AFM) calculated for $Cs_2AgF_4$ as a function of the Q distortion coordinate is displayed in Fig. 7. For simplicity, in that figure, E(FM) is chosen to be equal to zero for Q = 0. As a salient feature, Fig. 7 shows that a small orthorhombic distortion induces a change of sign of ΔE and thus $Cs_2AgF_4$ becomes FM at the final equilibrium geometry (Q = 0.34 Å). A similar situation is found[6] for $K_2CuF_4$. Experimental and calculated ΔE and J values at the equilibrium structures of $Cs_2AgF_4$ and $K_2CuF_4$ are collected in Table 6.

| Compound | Phase | ΔE(exper) | J(exper) | ΔE(calc) | J(calc) |
|---|---|---|---|---|---|
| $Cs_2AgF_4$ | Cmca | -7.58 | -3.79 | -11.2 | -5.6 |
| $K_2CuF_4$ | Cmca | -3.44 | -1.72 | -5.3 | -2.7 |

**Table 6.** Experimental and calculated ΔE and J values (in meV) at the equilibrium structures of $Cs_2AgF_4$ and $K_2CuF_4$ described in Tables 1 and 2. Experimental data come from [28] for $Cs_2AgF_4$ and [7] for $K_2CuF_4$.

It is worthwhile to remark that, on passing from Q = 0 to the equilibrium geometry in Fig. 7, there is an energy gain of around 140 meV per formula unit driven by the vibronic interaction of Eq. (8). When this quantity is compared to $\Delta E$ = 7.92 meV at Q = 0 (Table 4) we then realize that the magnetism in $Cs_2AgF_4$ is just a consequence of vibronic coupling in the electronic ground state.

The switch to a FM arrangement induced by the orthorhombic distortion in $Cs_2AgF_4$ can be understood through the ideas by Anderson[77,78] on a $M_A$ - F - $M_B$ dimer which in that situation is no longer symmetric. If we consider the complex associated with the cation at A (Fig. 2), then the charge transferred to ligands placed at *local x* axis is expected to be higher than that corresponding to ligands placed along the local *y* axis that involve a higher metal-ligand distance as a result of the orthorhombic distortion. Accordingly, we can write

$$f_\sigma(A,x) > f_\sigma(A,y)$$

$$f_\sigma(B,x) = f_\sigma(A,x) > f_\sigma(B,y) = f_\sigma(A,y) \qquad (21)$$

A central issue in the analysis of a non-symmetric dimer is the dependence of the J(AFM) contribution upon the quantities $f_\sigma(A,x)$ and $f_\sigma(B,y)$ involved in the bridging ligand of Fig. 2. Following a similar procedure to that by Anderson in the symmetric dimer[77] it can easily be obtained[6]

$$J(AFM) \propto f_\sigma(A,x) f_\sigma(B,y) = f_\sigma(A,x) f_\sigma(A,y) \qquad (22)$$

This result just means that, if bonding is significantly reduced in the longer F - $M_B$ bond, the exchange constant, J, could then be dominated by the FM contribution, J(FM), which is less sensitive to chemical bonding than J(AFM). Indeed the dependence of J(FM) on $f_\sigma(A,x)$ and $f_\sigma(A,y)$ has the form[6] $p_x f_\sigma(A,x) + p_y f_\sigma(A,y)$ thus implying that J(FM) is not zero even if $f_\sigma(A,y) = 0$.

According to Eq. (22), it is therefore crucial to determine the charge transferred to the ligand in the short, $f_\sigma(A,x)$, and long, $f_\sigma(A,y)$, bonds in the equilibrium geometry of both $Cs_2AgF_4$ and $K_2CuF_4$ compounds. Henceforth, these charges are simply denoted as $f_\sigma(x)$ and $f_\sigma(y)$.

Results of present calculations on this central issue are portrayed in Table 7. As a salient feature the obtained value of $f_\sigma(y)$ is equal to zero at the Cmca equilibrium geometry of both $Cs_2AgF_4$ and $K_2CuF_4$ compounds, a fact that strongly suggests that the AFM contribution, J(AFM), disappears. In other words, this key fact means that the orthorhombic distortion unveils the J(FM) contribution to J that is somewhat hidden in the I4/mmm parent phase. According to this view and the calculated values in Tables 4 and 5, we can estimate J(AFM) $\approx$ 9.6 meV and J(FM) $\approx$ 5.6 meV for the I4/mmm parent phase of $Cs_2AgF_4$, while J(AFM) $\approx$ 4.9 meV and J(FM) $\approx$ 2.7 meV would correspond to $K_2CuF_4$.

It is worthwhile to remark that the present results are underpinned by EPR data carried out on $CuCl_4X_2^{2-}$ centers (X = $NH_3$, $H_2O$) formed in $NH_4Cl:Cu^{2+}$ crystals grown in solutions with different pH[89,90,75]. While the equilibrium geometry of $CuCl_4(NH_3)_2^{2-}$ is tetragonal with four equivalent $Cl^-$ ions, lying in a plane perpendicular to the main axis, in $CuCl_4(H_2O)_2^{2-}$ there is an orthorhombic distortion in the plane formed by chlorine ions[70,75]. Interestingly, no superhyperfine interaction of the unpaired electron with the two far chlorine nuclei of the $CuCl_4(H_2O)_2^{2-}$ center have been observed by EPR[75], thus proving that $f_\sigma(y) = 0$. Moreover, from EPR data the calculated charge transferred to one of close Cl nuclei in $CuCl_4(H_2O)_2^{2-}$ ($f_\sigma(x)$ = 8.7%) is about twice that found for the tetragonal $CuCl_4(NH_3)_2^{2-}$ species as it is also shown on Table 7.

| System | Complex | Symmetry | $R_x$ | $R_y$ | $f_\sigma(x)$ (%) | $f_\sigma(y)$ (%) |
|---|---|---|---|---|---|---|
| $K_2CuF_4$ | $CuF_6^{4-}$ | Tetragonal | 2.061 | 2.061 | 1.70 | 1.70 |
| | | Orthorhombic | 1.941 | 2.234 | 4.86 | 0 |
| $Cs_2AgF_4$ | $AgF_6^{4-}$ | Tetragonal | 2.292 | 2.292 | 2.43 | 2.43 |
| | | Orthorhombic | 2.106 | 2.458 | 6.82 | 0 |
| $NH_4Cl:Cu^{2+}$ (pH=7) | $CuCl_4(NH_3)_2^{2-}$ | Tetragonal | 2.78 | 2.78 | 4.70 | 4.70 |
| $NH_4Cl:Cu^{2+}$ (pH=3) | $CuCl_4(H_2O)_2^{2-}$ | Orthorhombic | 2.52 | 2.97 | 8.7 | 0 |

**Table 7.** Charge transferred to ligands in the layer plane for the Cmca equilibrium geometry of $K_2CuF_4$ and $Cs_2AgF_4$. In this table $f_\sigma(x) \equiv f_\sigma(A,x)$ and $f_\sigma(y) \equiv f_\sigma(A,y)$ correspond to close and far ligands of a given complex lying along x and y directions, respectively. For the sake of completeness, the value $f_\sigma(x) = f_\sigma(y)$ associated with the tetragonal parent phase is also collected. The results are compared to those derived from EPR experiments on $CuCl_4(NH_3)_2^{2-}$ and $CuCl_4(H_2O)_2^{2-}$ centers formed in $NH_4Cl:Cu^{2+}$ crystals grown in solution with different pH. While the equilibrium geometry of $CuCl_4(NH_3)_2^{2-}$ is tetragonal with the four $Cl^-$ ions, lying in a plane perpendicular to the main axis, at the same distance from $Cu^{2+}$ in $CuCl_4(H_2O)_2^{2-}$ there is an orthorhombic distortion in the plane formed by chlorine ions. The values of $R_x$ and $R_y$ distances (in Å) for these centers are taken from [70].

The present results thus prove that the FM displayed in the layer planes of $Cs_2AgF_4$ and $K_2CuF_4$ is strongly dependent on the orthorhombic distortion[6]. As discussed in section 3.2 that distortion is fostered by the allowed vibronic coupling of the $^2A_{1g}$ ground state of a tetragonal $MF_6^{4-}$ complex (M = Cu, Ag) with the excited $^2B_{1g}$ state through the local $b_{1g}$ distortion mode.

It is worth noting now that a similar situation cannot take place for $K_2NiF_4$ where the ground state of the tetragonal $NiF_6^{4-}$ complex is $^3B_{1g}$ simply described[82] by the Slater determinant $|x^2-y^2\uparrow 3z^2-r^2\uparrow|$. Indeed, the $^1A_{1g}$ excited state $|3z^2-r^2\uparrow 3z^2-r^2\downarrow|$ cannot vibronically be coupled to the $^3B_{1g}$ ground state as the spin in both states is different, a matter already discussed in section 3.2. Furthermore, a $^3B_{1g}$ ground state can only vibronically be coupled to $^3A_{1g}$ excited states if a $b_{1g}$ distortion mode is involved. However, within the 45 states emerging from the $d^8$ configuration there are no $^3A_{1g}$ states in a local $D_{4h}$ symmetry[82]. This simple reasoning thus sheds some light on the lack of orthorhombic distortion in $K_2NiF_4$ and thus its AFM behavior. Along this line the ground state of the $MnF_6^{4-}$ complex in the I4/mmm $K_2MnF_4$ crystal has a spin S = 5/2. Accordingly, it cannot be coupled vibronically to any of the rest of 246 states coming from the $d^5$ configuration[82] which have at most S = 3/2. This relevant fact also helps to avoid the orthorhombic instability in $K_2MnF_4$. According to this reason, the local $O_h \rightarrow T_d$ symmetry change observed[91] for $BaF_2$:$Mn^{2+}$ involves the coupling of the electronic ground state (S = 5/2) with a charge transfer state of the same spin[92].

### 3.6. Insight into Rb$_2$AgF$_4$

The Rb$_2$AgF$_4$ compound was also firstly synthesized by Odenthal et al[25] in 1974 and subsequently explored by Friebel and Reinen[43] through EPR and optical spectroscopy. Despite the structure of Rb$_2$AgF$_4$ was not determined, the closeness of their experimental EPR and optical data to those reported for Cs$_2$AgF$_4$ already suggested that both compounds could display the same crystal structure. In 2016, Kurzydłowski et al. carried out DFT+U calculations on the Rb$_2$AgF$_4$ compound[93]. Although these authors find that the stable phase of Rb$_2$AgF$_4$ is described by the Cmca space group, they do not report the values of lattice parameters nor Ag$^{2+}$-F$^-$ distances.

|        |     | a      | b      | c      | β    | $R_z$ | $R_x$ | $R_y$ | $E_r$ |
|--------|-----|--------|--------|--------|------|-------|-------|-------|-------|
| Cmca   | FM  | 13.509 | 6.373  | 6.372  | 90   | 2.109 | 2.076 | 2.430 | 0     |
|        | AFM | 13.502 | 6.379  | 6.379  | 90   | 2.109 | 2.074 | 2.437 | +12   |
| I4/mmm | FM  | 4.521  | 4.521  | 13.360 | 90   | 2.066 | 2.261 | 2.261 | +150  |
|        | AFM | 4.522  | 4.522  | 13.350 | 90   | 2.063 | 2.261 | 2.261 | +133  |
| P2$_1$/c | FM  | 3.885  | 10.627 | 6.713  | 91.0 | 2.076 | 2.092 | 2.909 | +130  |
|        | AFM | 3.885  | 10.628 | 6.714  | 91.0 | 2.076 | 2.092 | 2.910 | +130  |

**Table 8.** Calculated lattice parameters (axes a, b, c, in Å, and monoclinic angle β, in degrees) and metal-ligand distances (axial $R_z$, and equatorials $R_x$, $R_y$, all in Å) for Rb$_2$AgF$_4$ in three different structures (orthorhombic Cmca, tetragonal I4/mmm and monoclinic P2$_1$/c) and two magnetic orders (FM, AFM). $E_r$ is the difference in energy (in meV/formula) with respect to the phase with minimum energy (FM Cmca). Note the different notation of the long axis in each structure.

For this reason, we have carried out DFT calculations, with the PW1PW hybrid functional, seeking to optimize the geometry of Rb$_2$AgF$_4$ in three possible layered phases namely Cmca, I4/mmm and P2$_1$/c. The P2$_1$/c structure is included as it seems to be involved[92] in $K_2AgF_4$ and $Na_2AgF_4$. Of course, for each kind of structure we have also explored both the FM and AFM arrangements of magnetic moments arising from silver ions. Results are gathered in Table 8 where the relative energy, $E_r$, of six considered phases is reported and $E_r = 0$ corresponds to the Cmca structure with FM ordering.

The calculated $E_r$ values in Table 8 mean that the stable phase in $Rb_2AgF_4$ corresponds to a Cmca structure with FM ordering such as it is experimentally found for $Cs_2AgF_4$. Moreover, if we consider for $Rb_2AgF_4$ an AFM ordering keeping the Cmca structure the changes of $Ag^{2+}$-$F^-$ distances are smaller than 0.01 Å and $E_r = 12$ meV. In this case $E_r$ is equal to the quantity $-\Delta E$ used in section 3.5. Thus, the value $E_r = -\Delta E = 12$ meV found for $Rb_2AgF_4$ is similar to that for $Cs_2AgF_4$ reflected in the results of Table 6 and Figure 7 of section 3.5. Along this line the values of $Ag^{2+}$-$F^-$ distances calculated for $Rb_2AgF_4$ in the stable phase (Table 8) are certainly close to $R_z = 2.112$ Å, $R_x = 2.111$ Å and $R_y = 2.441$ Å obtained for $Cs_2AgF_4$ (Table 1). This result is thus consistent with the practically identical d-d transitions[43] and EPR spectra[26,43] measured experimentally for both $Rb_2AgF_4$ and $Cs_2AgF_4$. For instance, the accurate EPR measurements carried out by Kurzydlowski et al. give[26] g(**a**) = 2.076 and g($\perp$**a**) = 2.26 for $Rb_2AgF_4$ and g(**a**) = 2.077 and g($\perp$**a**) = 2.257 for $Cs_2AgF_4$.

Concerning the I4/mmm parent phase with AFM ordering the present results on $Rb_2AgF_4$ yield $E_r = 133$ meV a quantity nearly identical to that derived for $Cs_2AgF_4$ such as it is shown in Figure 7 (section 3.5). According to the results of Table 8 such parent phase should be instable. We have verified that in that situation, there are two modes, $b_{1g}$ and $a_{2g}$, with the same imaginary frequency 401i cm$^{-1}$, which correspond to similar orthorhombic distortions of the F ligands within a plane but for the z = 0 and z = 1/2 layers, respectively.

As regards a possible monoclinic $P2_1/c$ structure, the results of Table 8 imply an energy per magnetic ion which is again 130 meV above that for the stable phase of $Rb_2AgF_4$.

**4. Final Remarks**

The present results stress the importance of exploring the I4/mmm parent phase of $K_2CuF_4$ and $Cs_2AgF_4$ for understanding the origin of their magnetic properties very distinct from those exhibited by other layered fluorides like $K_2NiF_4$ or $K_2MnF_4$. At the same time, they prove that the ferromagnetism observed for both $K_2CuF_4$ and $Cs_2AgF_4$ compounds greatly depends on the orthorhombic distortion[94]. As explained in sections 3.2 and 3.5 that instability is driven by the *allowed* vibronic admixture of the $^2A_{1g}$ ground state of a tetragonal $MF_6^{4-}$ complex (M = Cu, Ag) with the excited $^2B_{1g}$ state. On the contrary, the lack of excited states which, within the $d^5$ configuration, can be vibronically coupled with the ground state hamper the orthorhombic distortion in $K_2MnF_4$. This situation is thus similar to that taking place in $SrCl_2$ where the $d^9$ impurities $Ni^+$, $Cu^{2+}$ and $Ag^{2+}$ all move off-centre[21,22,95] while a $Mn^{2+}$ impurity remains[96] at the $Sr^{2+}$ site. Nevertheless, for $d^9$ impurities this situation changes when we replace $SrCl_2$ by $CaF_2$ where only Ni$^+$ moves off-center while Ag$^{2+}$ and Cu$^{2+}$ remain at the Ca$^{2+}$ site[22]. The substitution of chlorine by fluorine, the increase of the nominal charge of the cation and the reduction of interatomic distances all tend to increase the value of the K$_0$ contribution in Eq. (9) and thus they help to suppress the off-center instability.

These ideas also shed light on the occurrence of ferroelectricity in BaTiO$_3$ but not in CaTiO$_3$ involving smaller interatomic distances. Indeed, an *isolated* Ti$^{4+}$ cannot move spontaneously off-center in these perovskites but at equilibrium K$_0$ is higher for CaTiO$_3$ than for BaTiO$_3$ following the smaller interatomic distances. As ferroelectricity requires the *simultaneous* off-center motion of all Ti$^{4+}$ ions this explains[97], albeit qualitatively, why BaTiO$_3$ is better suited than CaTiO$_3$.

In the case of Cu$^{2+}$ complexes the existence of close excited states that can vibronically be mixed with the ground state favors changes of the electronic density and the associated distortion provided K < 0. This characteristic of Cu$^{2+}$ complexes not shared by high spin Mn$^{2+}$ or Ni$^{2+}$ ones is sometimes referred to as plasticity[98].

From the present analysis an understanding of optical and magnetic properties of layered compounds with d$^9$ ions requires to explore the influence of the electrostatic potential, $V_R(\mathbf{r})$, in the parent phase and the possible instabilities driven by a force constant that becomes negative. By contrast, the attempts to explain such properties assuming a JT effect in tetragonal, orthorhombic or monoclinic lattices are meaningless. An exception to this behavior is the widely studied AgF$_2$[99] whose parent phase is cubic. In this case the AgF$_6^{4-}$ units are trigonally distorted and thus under this symmetry the ground state is strictly degenerate making possible the existence of a JT effect as it has recently been discussed[50].

Further work on silver fluorides like K$_2$AgF$_4$[100] or Na$_2$AgF$_4$ is now underway.

## 5. Acknowledgments

We acknowledge financial support from Grant No. PGC2018-096955-B-C41 funded by MCIN/AEI/10.13039/501100011033. I. S.-M. (grant BDNS:589170) and G. S.-F. acknowledge financial support from Universidad de Cantabria and Gobierno de Cantabria.

## 9. TOC Graphic

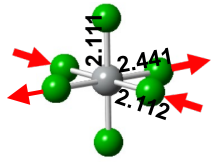

**Influence of orthorhombic distortion on magnetism**

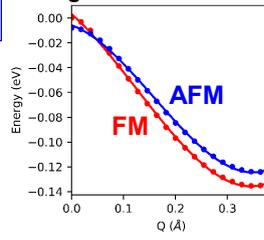

**Tetragonal → orthorhombic**